\documentclass[12pt]{article}
\usepackage{amsmath,amssymb,amsfonts,graphicx}

\begin{document}

\thispagestyle{empty}
\renewcommand{\refname}{References}

\title{ The Aharonov-Bohm effect in
scattering of short-wavelength particles}

\author{Yu. A. Sitenko$^1$ and N. D. Vlasii$^{1,2}$}

\date{}

\maketitle

\begin{center}
$^{1}$Bogolyubov Institute for Theoretical Physics, \\
National Academy of Sciences of Ukraine, \\ 14-b Metrologichna
Str., Kyiv, 03680, Ukraine \\
$^{2}$Physics Department, Taras Shevchenko National University of Kyiv, \\
64 Volodymyrska str., Kyiv, 01601, Ukraine
\end{center}

\begin{abstract}
Quantum-mechanical scattering of nonrelativistic charged particles
by a magnetic vortex of nonzero transverse size is considered. We
show that the flux of the vortex serves as a gate for the strictly
forward propagation of particles with short, as compared to the
transverse size of the vortex, wavelengths; this effect is the same
for a penetrable vortex as for an impenetrable one. A possibility
for the experimental detection of the scattering Aharonov-Bohm
effect is discussed.
\end{abstract}

PACS: 03.65.Ta, 75.70.Kw, 41.85.-p, 03.65.Nk

\bigskip

\begin{center}
Keywords: magnetic vortex, quantum-mechanical scattering, Fraunhofer
diffraction
\end{center}

\bigskip
\medskip

\section{Introduction}
The theoretical prediction of the Aharonov-Bohm (AB) effect in 1959
\cite{Aha} was one of the most intriguing achievements in quantum
theory. Now this effect has been long recognized for its crucial
role in demonstrating that, in addition to the usual local
(classical) influence of electromagnetic field on charged particles,
there exists the unusual nonlocal (purely quantum) influence of
electromagnetic fluxes confined in the regions which are
inaccessible to charged particles (see, {\it e.g.}, reviews
\cite{Pes,Ton}). A particular example is quantum-mechanical
scattering of nonrelativistic charged particles by an impenetrable
straight and infinitely long solenoid that encloses a magnetic flux:
as was shown in \cite{Aha}, this process depends periodically on the
value of the enclosed flux. Although a formula for the differential
cross section of this process was derived in Section 4. "Exact
solution for scattering problems" of \cite{Aha}, the formula was
never checked experimentally. Perhaps, it was not even intended to
be checked, since the authors \cite{Aha} in the preceding sections
of their work proposed a closely related, but different from the
scattering one, experiment to confirm their theoretical prediction;
this experiment consisted in observing a fringe shift in the
interference pattern due to two coherent particle beams under the
influence of an impenetrable magnetic vortex placed between the
beams. Since then the theory and the experiment for the AB effect
followed their own non-intersecting ways. The concern of
experimentalists was to exclude completely penetration of the
particle beams into the region of nonzero magnetic flux, see
\cite{Ton}. As to the theoretical development, scattering theory
initiated by Aharonov and Bohm was substantiated and further
elaborated, see \cite{Be,Rui,Jac}.

It should be noted that the concern of theoreticians was mostly in
the case of long-wavelength scattered particles, when the transverse
size of the magnetic vortex was neglected. Since a direct scattering
experiment is hard to perform with long-wavelength (slowly moving)
particles, thus elaborated theory remained actually unverified. On
the one hand, as the particle wavelength decreases, the perspectives
of performing a scattering experiment with such particles are
increasing. On the other hand, the transverse size of the vortex
then comes into play, and, given the technical difficulties of
measuring the interference AB effect, it may seem hardly possible to
avoid the particle penetration inside the vortex in a direct
scattering experiment. However, as we shall see, these misgivings
are irrelevant for the case of short-wavelength (fast-moving)
particles. The aim of the present paper is to extend scattering
theory to such a case (see also \cite{Si0,Si1,Si11}) and to reach
the realm where the experimental verification of the scattering AB
effect is quite feasible.

\section{Double-slit interference}
First, let us recall briefly the setup which is conventionally used
for the experimental verification of the AB effect (see, {\it e.g.},
\cite{Pes,Ton}). It involves the observation of the interference
patterns resulting from the two coherent electron beams bypassing
from different sides an impenetrable magnetic vortex which is
orthogonal to the plane defined by the beams. This is a so called
double-slit interference experiment, although in reality an
electrostatic biprism is used to bend the beams and to direct them
on the detection screen. Let the detection screen be parallel to the
screen with slits, $L$ be the distance between the screens, and $D$
be the distance between the slits. Otherwise, in the biprism
setting, the line connecting images of a source is parallel to the
detection screen, $L'$ is the distance between the line and the
screen, and $D'$ is the distance between the images; since the
interference pattern depends on quotient $D'/L'$ rather than on $L'$
and $D'$ separately, the primes will be dropped in the following.
The interference pattern on the detection screen consists of equally
spaced fringes which are in the same direction as the magnetic
vortex,
\begin{equation}
I(y)=4I_0(y)\cos^2\left[\left(\frac{yD}{\lambda L}+\frac{\Phi}{\Phi_0}\right)\pi\right],\label{eq1}
\end{equation}
where $y$ is the coordinate which is orthogonal to the fringes on
the detection screen ($y=0$ corresponds to the point which is
symmetric with respect to the slits), $I_0(y)$ is the intensity in
the case when either of the slits is closed, $\lambda$ is the
electron wavelength, $\Phi$ is the flux of the impenetrable magnetic
vortex placed just after the screen with slits (otherwise, after the
biprism), $\Phi_0=hce^{-1}$ is the London flux quantum. Intensity
$I(y)$ (1) is oscillating with period
\begin{equation}
\Delta=\lambda LD^{-1}\label{eq2},
\end{equation}
and the enveloping function is given by $4I_0(y)$ which is a
Gaussian centred at $y=0$. At the centre of the detection screen one
gets
\begin{equation}
I(0)=4I_0(0)\cos^2\left(\frac{\Phi}{\Phi_0} \pi\right).\label{eq3}
\end{equation}
If $L\gg D$ and $\lambda$, then one can use dimensionless (angular)
variable $\varphi=y/L$. The period of oscillations in this variable
is
\begin{equation}
\delta=\lambda D^{-1}.\label{eq4}
\end{equation}

Evidently, the period of oscillation decreases with the decrease of
wavelength $\lambda$. The linear resolution of the detector should
be at least as high as $ \frac 12 \Delta$, that is why the
observation of the interference pattern becomes more complicated in
the short-wavelength limit. Since the enveloping function takes a
form of a narrow peak in this limit, it is crucial that the
oscillations be distinguishable in the enveloping background. To
measure this, one defines the visibility of the central point as
\begin{equation}
V=\frac{\left|I(0)-I\left(\pm\frac 12\Delta\right)\right|}{I(0)+I\left(\pm\frac 12\Delta\right)}.\label{eq5}
\end{equation}
In view of the symmetry of the enveloping function
$\left(I_0\left(-\frac 12\Delta\right)=I_0\left(\frac
12\Delta\right)\right)$, one finds immediately
\begin{equation}
V=\frac{\left|I_0(0)\!-\!I_0\left(\!\frac 12\Delta\!\right)\!+\!
\left[I_0(0)\!+\!I_0\left(\!\frac 12\Delta\!\right)\right]\!\cos\!\left(\!2\frac{\Phi}{\Phi_0}\pi\!\right)\right|}
{I_0(0)\!+\!I_0\left(\frac 12\Delta\right)\!+\!\left[I_0(0)\!-\!I_0\left(\frac 12\Delta\right)\right]
\!\cos\!\left(2\frac{\Phi}{\Phi_0}\pi\right)}.\label{eq6}
\end{equation}
It should be noted that visibility (5) is nonzero in the case of the
absence of oscillations:
$V=|I_0(0)-I_0(\Delta/2)|/[I_0(0)+I_0(\Delta/2)]$. This case is
mimicked at $\Phi=(n\pm 1/4)\Phi_0$ and $\Phi=\left\{n\pm
\pi^{-1}\arcsin\left[I_0(0)/\sqrt{I_0^2(0)+I_0^2(\Delta/2)}\right]\right\}\Phi_0$
($n$ is an integer number).

Concluding this section, we note that the value of the bending
potential and the distance to the detection screen should be
adjusted in order that the interference pattern be visible. Such a
type of adjustment is unnecessary for the diffraction pattern in
direct scattering of short-wavelength particles on a magnetic
vortex.

\section{Direct scattering}
We start with the Schr\"{o}dinger equation for the wave function
describing the stationary scattering state,
\begin{equation}
H\psi(r,\,\varphi)=\frac{\hbar^2k^2}{2m}\psi(r,\,\varphi),\label{eq7}
\end{equation}
where $m$ is the particle mass and $k$ is the absolute value of the
particle wave vector ($k=2\pi/\lambda$); the impenetrable magnetic
vortex is assumed to be directed orthogonally to the plane with
polar coordinates $r$ and $\varphi$, and we confine ourselves to the
particle motion in this plane, since the motion along the vortex is
free. Out of the vortex core the Schr\"{o}dinger hamiltonian takes
the form
\begin{equation}
H=-\frac{\hbar^2}{2m}\left[r^{-1}\partial_rr\partial_r+r^{-2}\left(\partial_\varphi-
{\rm i}\Phi\Phi_0^{-1}\right)^2\right],\label{eq8}
\end{equation}
and we impose condition
\begin{equation}
\lim\limits_{r\rightarrow \infty}{\rm e}^{{\rm i}kr}\psi(r,\,\pm\pi)=1,\label{eq9}
\end{equation}
signifying that the incident wave comes from the far left; the
forward direction is $\varphi=0$, and the backward direction is
$\varphi=\pm \pi$.

Without a loss of generality we assume that the vortex has a shape
of cylinder of radius $r_c$ and impose the Robin boundary condition
on the wave function:
\begin{equation}
\left.\left\{\left[\cos(\rho\pi)+r_c\sin(\rho\pi)\partial_r\right]\psi(r,\,\varphi)\right\}\right|_{r=r_c}=0;\label{eq10}
\end{equation}
$\rho=0$ corresponds to the Dirichlet boundary condition (perfect
conductivity of the boundary), and $\rho=1/2$ corresponds to the
Neumann one (absolute rigidity of the boundary). The solution to (7)
with hamiltonian (8), which satisfies conditions (9) and (10), takes
the following form
\begin{equation}
\psi(r,\,\varphi)=\sum\limits_{n\in \mathbb{Z}}{\rm e}^{{\rm i}n\varphi}e^{{\rm i}(|n|-\frac 12|n-\mu|\pi)}
\left[J_{|n-\mu|}(kr)-\Upsilon_{|n-\mu|}^{(\rho)}(kr_c)H_{|n-\mu|}^{(1)}(kr)\right],\label{eq11}
\end{equation}
where $\mathbb{Z}$ is the set of integer numbers,
$\mu=\Phi\Phi_0^{-1}$, $J_\alpha(u)$ and $H_{\alpha}^{(1)}(u)$ are
the Bessel and the first-kind Hankel functions of order $\alpha$,
and
\begin{equation}
\Upsilon_\alpha^{(\rho)}(u)=\frac{J_\alpha(u)}{H_\alpha^{(1)}(u)}\,\frac{\cot(\rho\pi)+u\partial_u\ln J_\alpha(u)}
{\cot(\rho\pi)+u\partial_u\ln H_\alpha^{(1)}(u)}.\label{eq12}
\end{equation}
Thus, wave function (11) consists of two parts: the one which will
be denoted by $\psi_0(r,\,\varphi)$ is independent of $r_c$, and the
other one which will be denoted by $\psi_c(r,\,\varphi)$ is
dependent on $r_c$.

Taking the asymptotics of the first part at large distances from the
vortex, $kr\gg 1$, one can get (see \cite{Som})
\begin{eqnarray}
\psi_0(r,\,\varphi)={\rm e}^{{\rm i}kr\cos\varphi+{\rm i}\mu\varphi}\Bigl\{\cos(\mu\pi)-
{\rm i}{\rm sgn}(\varphi)\sin(\mu\pi)\Bigr.\nonumber\\
\Bigl.\times\left[1-{\rm e}^{{\rm i}\left(\frac 12+[\![\mu]\!]-\mu\right)\varphi}{\rm erfc}
\left({\rm e}^{-{\rm i\pi/4}}\sqrt{2kr}\left|\sin\frac{\varphi}{2}\right|\right)\right]\Bigr\},\label{eq13}
\end{eqnarray}
where $[\![u]\!]$ denotes the integer part of quantity $u$ ({\it
i.e.} the integer which is less than or equal to $u$), ${\rm
sgn}(u)=\left\{\begin{array}{cc}
                                           1, & u>0 \\
                                           -1, & u<0
                                         \end{array}\right\}$, ${\rm
erfc}(z)=\frac{2}{\sqrt{\pi}}\int\limits_{z}^{\infty}du\,{\rm
e}^{-u^2}$ is the complementary error function, and
it is implied that $-\pi<\varphi<\pi$. In the case
$\sqrt{kr}\left|\sin\frac{\varphi}{2} \right|\gg 1$, one obtains
\begin{equation}
\psi_0(r,\,\varphi)={\rm e}^{{\rm i}kr\cos\varphi}{\rm e}^{{\rm i}\mu[\varphi-{\rm sgn}(\varphi)\pi]}+
f_0(k,\,\varphi)\frac{{\rm e}^{{\rm i}(kr+\pi/4)}}{\sqrt{r}},\label{eq14}
\end{equation}
where
\begin{equation}
f_0(k,\,\varphi)={\rm i}\frac{{\rm e}^{{\rm i}\left([\![\mu]\!]+\frac 12\right)\varphi}}{\sqrt{2\pi k}}
\frac{\sin(\mu\pi)}{\sin(\varphi/2)}\label{eq15}
\end{equation}
is  the scattering amplitude which was first obtained by Aharonov
and Bohm \cite{Aha}. In the case $kr\gg 1$ but
$\sqrt{kr}\left|\sin\frac{\varphi}{2}\right|\ll 1$, one obtains
\begin{equation}
\psi_0(r,\,\varphi)={\rm e}^{{\rm i}kr\cos\varphi} \cos(\mu\pi).\label{eq16}
\end{equation}

Taking the large-distance asymptotics of the $r_c$-dependent part of
the wave function, one gets
\begin{equation}
\psi_c(r,\,\varphi)=f_c(k,\,\varphi)\frac{{\rm e}^{{\rm i}(kr+\pi/4)}}{\sqrt{r}},\label{eq17}
\end{equation}
where
\begin{equation}
f_c(k,\,\varphi)={\rm i}\sqrt{\frac{2}{\pi k}}\sum\limits_{n\in \mathbb{Z}}{\rm e}^{{\rm i}n\varphi}
{\rm e}^{{\rm i}(|n|-|n-\mu|\pi)}\Upsilon_{|n-\mu|}^{(\rho)}(kr_c).\label{eq18}
\end{equation}
In the long-wavelength limit, $kr_c\ll 1$, amplitude $f_c$ (18) is
suppressed by powers of $kr_c$ as compared to amplitude $f_0$ (15);
however, as was already mentioned, this limit is not feasible to
experimental measurements. In the short-wavelength limit, $kr_c\gg
1$, amplitude $f_0$ (15) is suppressed and wave function $\psi_0$
(13) is actually reduced to a plane wave, ${\rm e}^{{\rm
i}kr\cos\varphi}$, which is distorted by the flux-dependent factors,
see (16) and the first term in (14). Amplitude $f_c$ (18) in this
limit takes form \cite{Si11}
\begin{eqnarray}
&f_c(k,\,\varphi)={\rm i}\sqrt{\frac{2\pi}{k}}\left[\cos(\mu\pi)
\Delta^{([\![\mu]\!])}_{kr_c}(\varphi)-\sin(\mu\pi)\Gamma^{([\![\mu]\!])}_{kr_c}(\varphi)\right]\nonumber \\
&-\sqrt{\frac{r_c}{2}|\sin(\varphi/2)|}\exp\left\{-2{\rm i}kr_c|\sin(\varphi/2)| +{\rm
i}\mu[\varphi-{\rm sgn}(\varphi)\pi]\right\} \nonumber \\ &\times\exp\left\{-{\rm
i}\left[2\chi^{(\rho)}(kr_c,\varphi)+\pi/4\right]\right\}+\sqrt{r_c}O\left[(kr_c)^{-1/6}\right],
\,\,kr_c\gg 1,\label{eq19}
\end{eqnarray}
where
\begin{equation}
\Delta_x^{([\![\mu]\!])}(\varphi)=\frac{1}{2\pi}\sum\limits_{|n-\mu|\leq x}{\rm e}^{in\varphi},\,\,\,\,\,\,
\Gamma_x^{([\![\mu]\!])}(\varphi)=\frac{1}{2\pi{\rm i}}\sum\limits_{|n-\mu|\leq x}{\rm sgn}(n-\mu){\rm e}^
{{\rm i}n\varphi},\label{eq20}
\end{equation}
and
\begin{equation}
\chi^{(\rho)}(kr_c,\varphi)={\rm arctan}\left[\frac{2kr_c|\sin^3(\varphi/2)|}
{2{\rm cot}(\rho\pi)\sin^2(\varphi/2)-1}\right].\label{eq21}
\end{equation}
This amplitude satisfies the optical theorem of the unusual form due
to the long-range nature of the interaction with a vortex
\cite{Si0}:
\begin{equation}
2\sqrt{\frac{2\pi}{k}}\cos(\mu\pi){\rm Im}f_c(k,\,0)+4r_c\sin^2(\mu\pi)=
\int\limits_{-\pi}^{\pi}{\rm d}\varphi|f_c(k,\,\varphi)|^2.\label{eq22}
\end{equation}
The differential cross section\footnote{It is a cross section per
unit length along the vortex axis, hence its dimension is that of
length.} in the short-wavelength limit is given by expression
\begin{eqnarray}
\frac{{\rm d}\sigma}{{\rm d}\varphi}\equiv |f_c(k,\varphi)|^2=2r_c\biggl\{\cos(2\mu\pi){\Delta}_{kr_c}(\varphi)
\biggr.\nonumber\\ \biggl.+\!\left[1\!-\!\cos(2\mu\pi)\!-\!\sin(2\mu\pi)\sin(kr_c\varphi)\right]
{\Delta}_{\frac12kr_c}(\varphi)\!\biggr\}\!+\!\frac{r_c}{2}\left|\sin\frac\varphi 2\right|,\label{eq23}
\end{eqnarray}
where
\begin{equation}
\Delta_x(\varphi)=\frac{1}{4\pi x}\frac{\sin^2(x\varphi)}{\sin^2(\varphi/2)}\qquad(-\pi<\varphi<\pi)\label{eq24}
\end{equation}
is a strongly peaked at $\varphi=0$ and $x\gg 1$ function which can
be regarded as a regularization of the angular delta-function,
\begin{equation}
\lim\limits_{x\rightarrow\infty}\Delta_x(\varphi)=\frac{1}{2\pi}\sum\limits_{n\in \mathbb{Z}}{\rm e}^{{\rm i}n\varphi},
\,\,\,\,\,\,
\Delta_x(0)=\frac{x}{\pi}.\label{eq25}
\end{equation}
The first term on the right-hard side of (23) describes the forward
peak of the Fraunhofer diffraction on the vortex, while the last
term describes the classical reflection from the vortex according to
the laws of geometric (ray) optics. It should be noted that the
scattering amplitude depends on the choice of a boundary condition
via phase factor $\exp(-2{\rm i}\chi^{(\rho)})$, see (19) and (21);
therefore, the differential cross section is independent of boundary
conditions at all.

Moreover, the Fraunhofer diffraction is not affected by the
penetrability of the magnetic vortex, when a boundary condition at
the edge of the vortex is changed to a matching condition there. In
the case of an arbitrary cylindrically symmetric configuration of
the magnetic field strength inside the vortex of finite flux, the
wave function in the interior of the vortex can be presented as
\begin{equation}
\kappa(r,\,\varphi)=\sum\limits_{n\in \mathbb{Z}}{\rm e}^{{\rm i}n\varphi}
c_n^{(k)}\kappa_n(kr)\qquad (r<r_c),\label{eq26}
\end{equation}
where $\kappa_n(kr)$ is a regular solution to the appropriate
partial wave equation. Matching the logarithmic derivatives of
partial radial components of the solutions at the edge of the
vortex, one can find that the wave function in the exterior of the
vortex takes the form of (11) with
$\Upsilon_{|n-\mu|}^{(\rho)}(kr_c)$ substituted by
$\Upsilon_n(kr_c)$, where
\begin{equation}
\Upsilon_n(u)=\frac{J_{|n-\mu|}(u)}{H_{|n-\mu|}^{(1)}(u)}
\frac{\partial_u\ln \kappa_n(u)-\partial_u\ln J_{|n-\mu|}(u)}
{\partial_u\ln \kappa_n(u)-\partial_u\ln H_{|n-\mu|}^{(1)}(u)}.\label{eq27}
\end{equation}
Hence, the $r_c$-dependent part of the scattering amplitude is, cf.
(18),
\begin{eqnarray}
f_c(k,\,\varphi)={\rm i}\sqrt{\frac{2}{\pi k}}\sum\limits_{n\in\mathbb{Z}}{\rm e}^{{\rm i}n\varphi}
{\rm e}^{{\rm i}(|n|-|n-\mu|)\pi}\Upsilon_n(kr_c)= \nonumber \\
=f_c^{(\rm peak)}(k,\,\varphi)+f_c^{(\rm class)}(k,\,\varphi)+f_c^{(\rm res)}(k,\,\varphi),\label{eq28}
\end{eqnarray}
where
\begin{equation}
f_c^{(\rm peak)}(k,\,\varphi)=\frac{\rm i}{\sqrt{2\pi k}}\sum\limits_{|n-\mu|\leq kr_c}
{\rm e}^{{\rm i}n\varphi}{\rm e}^{{\rm i}(|n|-|n-\mu|)\pi},\label{eq29}
\end{equation}
\begin{equation}
f_c^{(\rm class)}(k,\,\varphi)=\frac{\rm i}{\sqrt{2\pi k}}\sum\limits_{|n-\mu|\leq kr_c}
{\rm e}^{{\rm i}n\varphi}{\rm e}^{{\rm i}(|n|-|n-\mu|)\pi}\tilde{\Upsilon}_n(kr_c),\label{eq30}
\end{equation}
\begin{equation}
f_c^{(\rm res)}(k,\,\varphi)={\rm i}\sqrt{\frac{2}{\pi k}}\sum\limits_{|n-\mu|> kr_c}
{\rm e}^{{\rm i}n\varphi}{\rm e}^{{\rm i}(|n|-|n-\mu|)\pi}\Upsilon_n(kr_c),\label{eq31}
\end{equation}
and
\begin{equation}
\tilde{\Upsilon}_n(u)=\frac{H^{(2)}_{|n-\mu|}(u)}{H^{(1)}_{|n-\mu|}(u)}
\frac{\partial_u\ln \kappa_n(u)-\partial_u\ln H^{(2)}_{|n-\mu|}(u)}
{\partial_u\ln \kappa_n(u)-\partial_u\ln H^{(1)}_{|n-\mu|}(u)};\label{eq32}
\end{equation}
$H^{(2)}_\alpha(u)=\left[H^{(1)}_\alpha(u)\right]^*$is the
second-kind Hankel function of order $\alpha$.

The finite sum of geometric progression, see (29), is independent of
the matching condition and can be rewritten as
\begin{equation}
f^{(\rm peak)}_c(k,\,\varphi)={\rm i}\sqrt{\frac{2\pi}{k}}
\left[\cos(\mu\pi)\Delta^{([\![\mu]\!])}_{kr_c}(\varphi)-\sin(\mu\pi)
\Gamma^{([\![\mu]\!])}_{kr_c}(\varphi)\right].\label{eq33}
\end{equation}
This is the amplitude of the Fraunhofer diffraction which is exactly
the same as in the case of the impenetrable vortex, see the first
term on the right-hand side of (19) (the explicit form of
$\Delta^{([\![\mu]\!])}_x(\varphi)$ and
$\Gamma_x^{([\![\mu]\!])}(\varphi)$ (20) is given in \cite{Si11}).

The residual infinite sum, see (31), is estimated at $kr_c\gg 1$ in
a similar way as in Appendix A of \cite{Si11}, yielding
\begin{equation}
f_c^{(\rm res)}(k,\,\varphi)=\sqrt{r_c}O[(kr_c)^{-1/6}].\label{eq34}
\end{equation}

Amplitude (30) depends on the matching condition and, consequently,
on the configuration of the magnetic field strength inside the
vortex. In the limit of short wavelengths, $kr_c\gg 1$, the squared
absolute value of the amplitude yields the differential cross
section of the classical scattering off the penetrable vortex.
However, there are two features which are quite general for this
scattering: the estimate of the amplitude in the strictly forward
direction is
\begin{equation}
f_c^{(\rm class)}(k,\,0)=\sqrt{r_c}O[(kr_c)^{-1/2}],\label{eq35}
\end{equation}
and the integrated cross section is
\begin{equation}
\sigma^{(\rm class)}\equiv\int\limits_{-\pi}^{\pi}{\rm d}\varphi\left|
f_c^{(\rm class)}(k,\,\varphi)\right|^2=2r_c.\label{eq36}
\end{equation}

\section{Fringe shift in the diffraction pattern}
Using (23), with the notations $d=2r_c$ (for the vortex diameter)
and $\lambda=2\pi/k$ (for the particle wavelength), as well as the
relation
$$
\Delta_x(\varphi)=2\Delta_{\frac 12x}(\varphi)\cos^2\left(\frac 12x\varphi\right),
$$
we present the differential cross section for scattering of
short-wavelength particles in a form similar to (1):
\begin{equation}
\frac{\rm d\sigma}{\rm d\varphi}=2d\Delta_{\frac{d\pi}{2\lambda}}(\varphi)\cos^2
\left[\left(\frac{\varphi d}{2\lambda}+\frac{\Phi}{\Phi_0}\right)\pi\right]+
\frac{\rm d\sigma^{(\rm class)}}{\rm d\varphi},\label{eq37}
\end{equation}
where, in the case of the impenetrable vortex, one has
\begin{equation}
\frac{\rm d\sigma^{(\rm class)}}{\rm d\varphi}=\frac{d}{4}\left|\sin\frac{\varphi}{2}
\right|.\label{eq38}
\end{equation}
The differential cross section of the Fraunhofer diffraction is
oscillating with period, cf. (4),
\begin{equation}
\delta=2\lambda d^{-1},\label{eq39}
\end{equation}
and the enveloping function is
$2d\Delta_{\frac{d\pi}{2\lambda}}(\varphi)$. In the strictly forward
direction one gets, cf. (3),
\begin{equation}
\left.\frac{\rm d\sigma}{\rm d\varphi}\right|_{\varphi=0}=\frac{d^2}{\lambda}\cos^2\left(
\frac{\Phi}{\Phi_0}\pi\right).\label{eq40}
\end{equation}
Assuming that the angular resolution of the detector is $\frac
12\delta$, we define the visibility of the central point in the
differential cross section as
\begin{equation}
V=\frac{\left|\rm d\sigma|_{\varphi=0}-\rm d\sigma|_{\varphi=\pm \frac 12\delta}\right|}
{\rm d\sigma|_{\varphi=0}+\rm d\sigma|_{\varphi=\pm\frac 12\delta}},\label{eq41}
\end{equation}
and obtain immediately
\begin{equation}
V=\frac{\left|1-\frac{4}{\pi^2}+\left(1+\frac{4}{\pi^2}\right)\cos\left(2\frac{\Phi}{\Phi_0}\pi\right)\right|}
{1+\frac{4}{\pi^2}+\left(1-\frac{4}{\pi^2}\right)\cos\left(2\frac{\Phi}{\Phi_0}\pi\right)}.\label{eq42}
\end{equation}
The maximal visibility ($V=1$) is attained for the flux which is
quantized in the units of the Abrikosov vortex flux
$\left(\Phi=\frac n2\Phi_0\right)$; the minimal visibility ($V=0$)
is attained at $\Phi=\Phi_{n\,\pm}$, where
\begin{equation}
\Phi_{n\,\pm}=\left[n\pm\frac 14\pm\frac{1}{2\pi}\arcsin\left(\frac{1-4/\pi^2}{1+4/\pi^2}\right)\right]\Phi_0.\label{eq43}
\end{equation}

The dependence on the vortex flux is washed off after integration
over the angle, and the contribution of the Fraunhofer diffraction
to the total cross section is equal to that of the classical
scattering (36). Thus, the total cross section is
\begin{equation}
\sigma_{\rm tot}=2d,\label{eq44}
\end{equation}
that is twice the classical total cross section. The optical theorem
relates the total cross section which is the right-hand side of (22)
to the scattering amplitude in the strictly forward direction, which
stands on the left-hand side of (22); it should be emphasized that
only the Fraunhofer diffraction contributes to the latter, whereas
both the Fraunhofer diffraction and the classical scattering
contribute to the former.

\begin{figure}
\includegraphics[width=250pt]{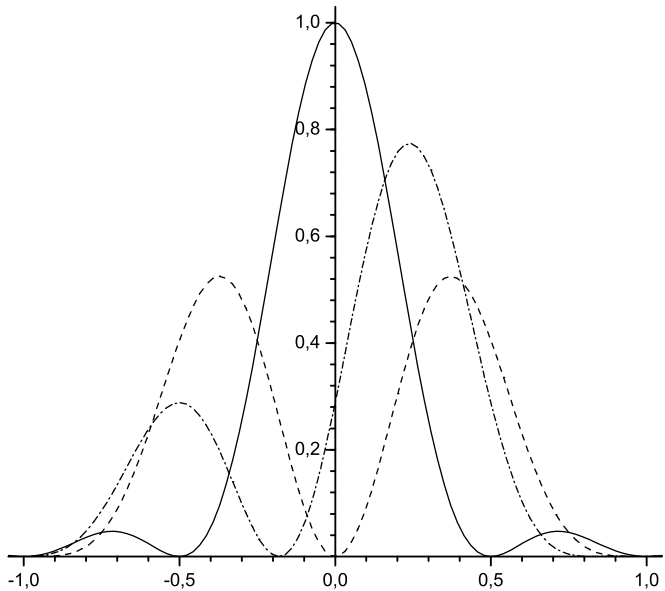}\\
\caption{$\delta{\rm d\sigma}/(\sigma_{\rm
tot}{\rm d\varphi})$ is along the ordinate axis and $\varphi/\delta$
is along the abscissa axis. Solid ($\Phi=\Phi_0$) and dashed
($\Phi=\frac 12\Phi_0$) lines correspond to $V=1$, whereas a dot-and-dash
($\Phi=\Phi_{1\,-}$) line corresponds to $V=0$.}\label{1}
\end{figure}\normalsize

The oscillations of the differential cross section in the central
region of the forward direction, $-\delta<\varphi<\delta$, are
depicted on fig.1 in the cases of maximal and minimal visibilities;
the pattern is almost the same for a wide range of short
wavelengths, $10^2<d\lambda^{-1}<10^6$. The area under plotted lines
is 0.4749701$\pm$0.0000004 for $\Phi=\Phi_0$,
0.4414430$\pm$0.0000010 for $\Phi=\Phi_{1\,-}$ and
0.4278549$\pm$0.0000013 for $\Phi=\frac 12\Phi_0$. Thus, the
contribution of oscillations from the outer region,
$\delta<|\varphi|<\pi$, is less than 17\% of that from the central
one; it is even less than 6\% in the case $\Phi=\Phi_0$. The area
under the one central peak in the case $\Phi=\Phi_0$ ($-\frac
12\delta<\varphi<\frac 12\delta$) is 0.4514119$\pm$0.0000002; it
exceeds the area under the two peaks in the cases $\Phi=\Phi_{1\,-}$
and $\Phi=\frac 12\Phi_0$ ($-\delta<\varphi<\delta$). The
contribution of the classical scattering to the central region is at
all negligible.

\section{Summary and discussion}
We have shown that the fringe shift emerging under the influence of
a magnetic vortex in the diffraction pattern in a direct scattering
experiment with short-wavelength particles, see (37), is completely
analogous to that in the interference pattern in a double-slit
experiment, see (1). However, there are some features which are
peculiar to the case of short-wavelength particles.

During the 50 years long history of the experimental verification of
the AB effect (see \cite{Ton}), many efforts were undertaken to
ensure the impenetrability of the magnetic vortex, since the issue
of penetration of particle beams into the region of the magnetic
flux was the principal one to seed doubts in the verification.
Perhaps, these misgivings had grounds in the case of long-wavelength
particles, but, as follows from scattering theory, there is no room
for them in the case of short-wavelength particles. The
penetrability of the magnetic vortex does not affect the diffraction
pattern (first term in (37)), only the classical reflection (given
by (38)) is affected. It may seem to be somewhat paradoxical, but
the AB effect with short-wavelength particles is the same for a
penetrable vortex as for an impenetrable one.

The overwhelming contribution to the whole diffraction cross section
comes from a narrow region, $-\delta<\varphi<\delta$, around the
strictly forward direction. A gate for the strictly forward
propagation of short-wavelength particles is opened for the magnetic
flux equal to the flux of even number of the Abrikosov vortices,
say, at $\Phi=\Phi_0$: more than 90\% of the diffraction cross
section is given by a peak centred at $\varphi=0$ and having width
$\delta$. As the flux diminishes, the peak is lowered and shifted to
the right. If the angular resolution is $\frac 12\delta$, then the
visibility of the central point is maximal ($V=1$) at $\Phi=\Phi_0$,
diminishing with the diminishing flux and achieving its minimum
($V=0$) at $\Phi=\Phi_{1\,-}$, see (43). Furthermore, the visibility
is growing with the diminishing flux and achieves its maximum
($V=1$) for the flux equal to that of the Abrikosov vortex, at
$\Phi=\frac 12\Phi_0$ (or odd number of the Abrikosov vortices in
general). The pattern is again symmetric but with a dip in the
strictly forward direction: the gate is closed. This gate effect is
illustrated by fig.1 for a wide range of short wavelengths,
$10^2<d\lambda^{-1}<10^6$.

Certainly, the Fraunhofer diffraction ({\it i.e.} the diffraction in
almost parallel rays) is a well-known phenomenon of wave optics.
Poisson was the first who predicted theoretically in 1818 a spot of
brightness in the centre of a shadow of an opaque disc; the
prediction was in a contradiction with the laws of geometric (ray)
optics. It is curious that Poisson used his prediction as an
argument to disprove wave optics which had been just developed by
Fresnel: this demonstrates how unexpected and unbelievable Poisson's
result was at that time. Nevertheless, the brightness spot in the
centre of the disc shadow was observed; the decisive experiments
were performed by Arago and Fresnel. According to Sommerfeld
\cite{Somm}, the diffraction on the opaque disc bears the name of
Poisson and the brightness spot in the shadow centre bears the name
of Arago. The same effect persists for scattering of light on an
opaque sphere and other obstacles. However, in the case of obstacles
in the form of a long strip or cylinder, the streak of brightness in
the centre of a shadow of such obstacles might be elusive to the
experimental detection: as is noted in the eminent treatise
\cite{Mor}, it seems more likely that the measurable quantity is the
classical cross section, although the details of this phenomenon
depend on the method of measurement.

Almost six decades have passed from the time when this assertion was
made by Morse and Feshbach, and experimental facilities have
improved enormously since then. For instance, in optics, a streak of
brightness in the shadow of a hair can be observed with the use of
laser beams. In the present paper, we point at the circumstances
when the detection of the forward diffraction peak in electron
optics will be the detection of the AB effect as well. We propose to
perform a scattering experiment using electrons with the wavelength
of order 0.1 nm and a magnetic vortex (tiny ferromagnet or solenoid)
with the diameter of order 1 $\mu$m; a distance from the vortex to
the detector and that from the electron source to the vortex might
be of order 100 mm.

\section*{Acknowledgments}

The work was supported in part by the Ukrainian-Russian SFFR-RFBR
project F40.2/108 "Application of string theory and field theory
methods to nonlinear phenomena in low-dimensional systems".

\end{document}